# Enhancement of interlayer exciton emission in a TMDC heterostructure via a multi-resonant chirped microresonator up to room temperature

Chirag C. Palekar[1 *], Barbara Rosa[1], Niels Heermeier[1], Ching-Wen Shih[1], Imad Limame[1],

Aris Koulas-Simos[1], Arash Rahimi-Iman[2], Stephan Reitzenstein[1 **]

[1]Institut für Festkörperphysik, Technische Universität Berlin, Hardenbergstraße 36, 10623 Berlin, Germany

[2]I. Physikalisches Institut and Center for Materials Research, Justus-Liebig-Universität Gießen, Heinrich-Buff-Ring 16, 35392 Gießen, Germany

*c.palekar@tu.berlin.de; ** stephan.reitzenstein@physik.tu-berlin.de

## Abstract:

We report on multi-resonance chirped distributed Bragg reflector (DBR) microcavities. These systems are employed to investigate the light-mater interaction with both intra- and inter-layer excitons of transition metal dichalcogenide (TMDC) bilayer heterostructures. The chirped DBRs consisting of $SiO_2$ and $Si_3N_4$ layers with gradually changing thickness exhibit a broad stopband with a width exceeding 600 nm. Importantly, and in contrast to conventional single-resonance microcavities, our structures provide multiple resonances across a broad spectral range, which can be matched to spectrally distinct resonances of the embedded TMDC heterostructures. We study cavity-coupled emission of both intra- and inter-layer excitons from an integrated $WSe_2$/$MoSe_2$ heterostructure in a chirped microcavity system. We observe an enhanced interlayer exciton emission with a Purcell factor of 6.67 ± 1.02 at 4 K. Additionally, we take advantage of cavity-enhanced emission of the interlayer exciton to investigate its temperature-dependent luminescence lifetime, which yields a value of 60 ps at room temperature. Our approach provides an intriguing platform for future studies of energetically distant and confined excitons in different semiconducting materials, which paves the way for various applications such as microlasers and single-photon sources by enabling precise control and manipulation of excitonic interactions utilizing multimode resonance light-matter interaction.

## Introduction

Transition metal dichalcogenide (TMDC) monolayers (ML) are highly suited platforms for exploring the physics of 2D quantum materials and opto-electronic applications[1]. These materials interact effectively with light fields due to their high oscillator strength, high quantum efficiency, and large exciton binding energies[2–5]. Furthermore, van der Waals homo- and hetero-structures created by stacking TMDC MLs exhibit interlayer excitons possessing fascinating optical characteristics, including spin-valley polarized emission, and engineered lifetimes[6–10]. Of particular interest are twisted TMDC heterostructures (HS), in which the confinement potential creates a periodic trapping of excitons, forming an array of quantum-dot-like emission sites[11–14]. Trapped and interacting excitons in such arrays provide an attractive platform for implementing quantum information processing as well as for studying many-body interactions, such as correlated quantum states[15], Mott transitions[16], and superradiant emission effects[17].

The TMDC HSs exhibit multiple excitonic resonances such as intralayer $(X)$ and interlayer $(IX)$ excitons. The $X$ with electron and hole being in the same monolayer has high binding energy with stable and bright emission at room temperature (RT). On the other hand, spatially indirect $IX$ with their peculiarities bear some challenges regarding practical applications besides their attractive properties. For instance, due to electrons and holes residing in different layers, they typically show lower oscillator strengths and lower RT-emission yield than their intralayer counterparts[18–20]. The low light-matter coupling strength is attributed to the twist angle dependent interlayer separation, and to the momentum mismatch along with varying overlap of the electron-hole wavefunction[18]. This is understood as the reason behind the many orders of magnitude lower RT IX emission[20]. Therefore, external factors can help to address a comparably low emission rate, for example applying vertical pressure[21,22], strain engineering[23], and boosting the optical gain by more complex stacking configurations, such as trilayer configurations[24–27]. In addition, utilizing the Purcell effect[28] must be considered to enhance the emission from excitons in TMDC HSs with the goal of integration in photonic applications[5,29–32]. Considering conventional microcavity configurations based on top and bottom distributed Bragg reflectors (DBRs) with finite stopband width, it is a challenging task to couple and consequently, enhance the emission of all distinct and spectrally separated exciton resonances. Thus, a versatile and easily scalable microcavity configuration exhibiting capabilities of multi-resonance formation at relatively small effective cavity lengths is the key to generate the desired multi-mode light matter interaction.

Here we introduce an advanced microcavity configuration that allows for multi-resonance light-matter coupling of energetically separated exciton resonances of the integrated active medium. This approach is based on broadband chirped distributed Bragg reflectors (cDBRs) with a stopband width that covers a wide spectral range exciding 600 nm. The cDBR is composed of $SiO_2$ and $Si_3N_4$ layers, which are deposited on a Si substrate with a gradually changing thickness in a chirped manner. A fully fabricated chirped microcavity, consisting of a lower and an upper cDBR, features a variety of tailorable cavity modes that can be tuned within its broad stopband by design to enable resonant coupling with multiple energetically separated matter excitations for multi-resonant coupling applications. The layer thicknesses of the chirped microcavity are designed using the transfer matrix method, which leads to excellent agreement between the calculated and measured spectral positions of the cavity modes. Additionally, we experimentally demonstrate multi-resonance coupling between cavity modes and TMDC HS exciton resonances by integrating the twisted $WSe_2/MoSe_2$ HS inside a cDBR microcavity. The measured photoluminescence (PL) response shows emission from $X^{WSe_2}$, $X^{MoSe_2}$ and $IX$ coupled to the established microcavity modes. Moreover, we observe enhanced $IX$ emission at RT as required for target applications such as TMDC HS microlasers[33]. The enhanced emission at RT allows us so far uniquely to perform the temperature dependent lifetime measurements to get important insights into the underlying emission processes. We report shortened cavity-coupled interlayer lifetime due to the Purcell effect. Our approach goes beyond ordinary microcavity structure fabrication for light-matter interaction studies, providing an intricate avenue to investigate multi-resonant weak and strong coupling scenarios involving various active media integrated in the microcavities that possess energetically separated resonances.

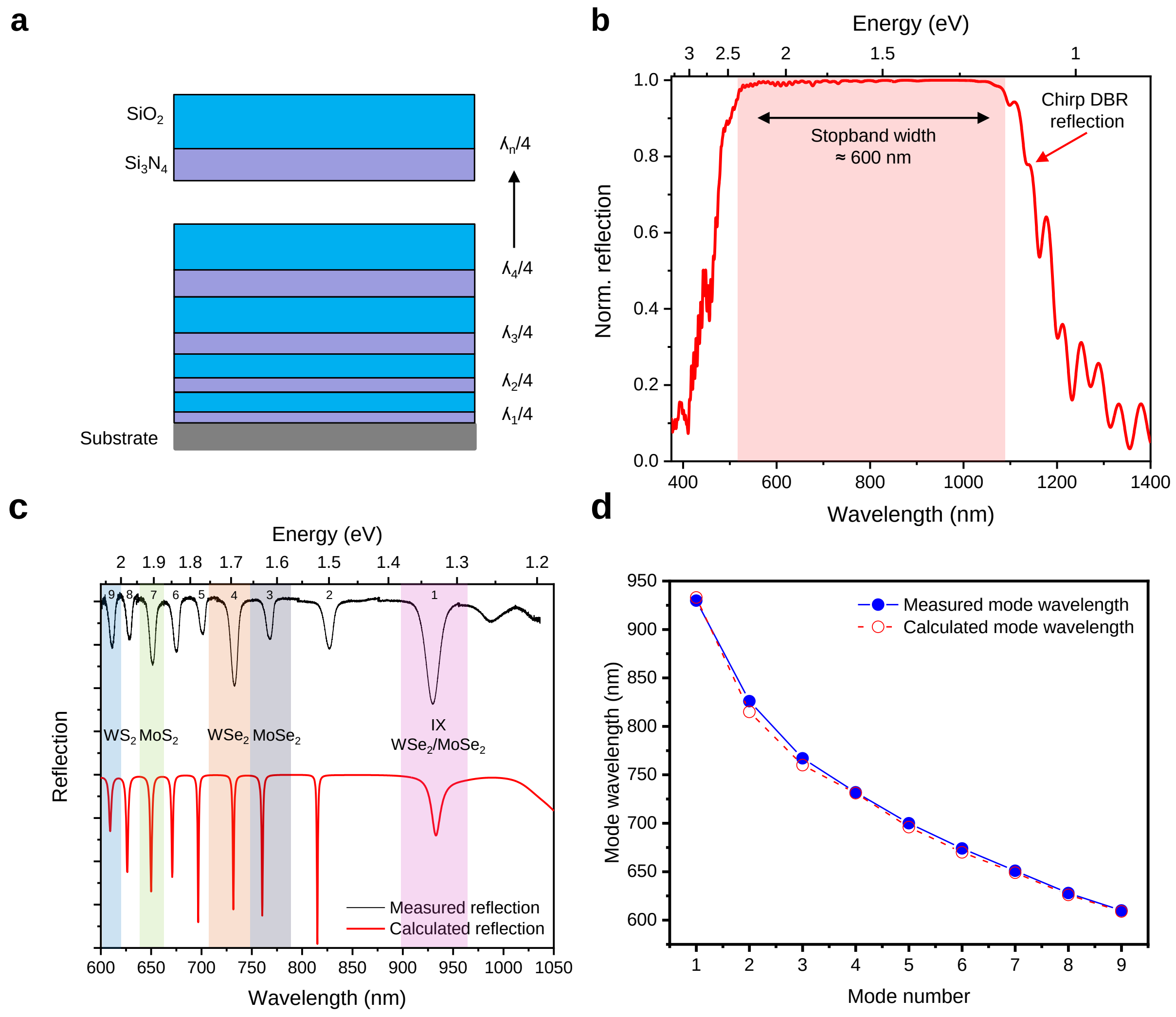

***Figure 1: Layer design and calculated reflection spectrum of a broadband chirped DBR microcavity. a)*** *Schematic side view of the designed cDBR. The DBR is composed of $SiO_2$ and $Si_3N_4$ layer pairs with a thickness gradient based on varying design wavelengths* $(\lambda_1, \lambda_2, \lambda_3, \dots, \lambda_n)$. ***b)*** *Calculated reflection response of the cDBR showcasing a broad stopband width of ≈ 600 nm, with stepwise increased $\lambda$ by 50 nm from 500 to 1050 nm. For each $\lambda$ step, four layer pairs are used, resulting in a total of 36 cDBR layer pairs (cf. Table S1.).* ***c)*** *Calculated (red) and measured (black) reflection spectrum of a chirped microcavity structure based on design parameters shown in Table 1. The color-highlighted regions indicate spectral positions of neutral exciton emission from semiconducting TMDC MLs of $WS_2$, $MoS_2$, $WSe_2$, and $MoSe_2$, as well as the $IX$ from $WSe_2/MoSe_2$ heterostructures, at room temperature. The cavity modes are numbered 1 to 9 from longer to shorter wavelength.* ***d)*** *Comparison between the extracted mode wavelengths from calculated (open circles) and measured (filled circles) reflection spectra (shown in panel c)) for mode numbers 1 to 9.*

## Broadband chirped distributed Bragg reflector design

Conventional DBRs are composed of alternating material layers with varying refractive indices. The thickness of each layer is fixed and given by $i \times \lambda_0/(4\, n_{mat})$, where *i* represents odd integer numbers, $\lambda_0$ is the design wavelength in vacuum, and $n_{mat}$ is the refractive index of the material. For simplicity, we refer to $(\lambda_0/\, n_{mat})$ as $\lambda_0$ from now on. In such conventional DBRs, $\lambda_0$ remains constant for all layers, which restricts the DBR stopband width. However, in our approach, we fabricated the DBRs with a broad stopband using high refractive index contrast materials such as $SiO_2$ and $Si_3N_4$ with a systematic thickness variation of $\lambda_0$ throughout the layers. We begin with the deposition of several layers of both materials, considering $\lambda_1$, then deposit a similar number of layers for $\lambda_2$ directly on top, and so forth up to the n$^{th}$ layer. Figure 1a illustrates the schematic side view of such a DBR stack. Repeating this process with a consistent design wavelength separation ($\lambda_1$-$\lambda_2$) in a chirped manner results in a DBR with a spectrally broad stopband, hence referred to as cDBR.

The cDBR consists of $SiO_2$ and $Si_3N_4$ layer pairs with refractive index of 1.470 and 2.011 (at RT), respectively. The stopband width of conventional DBR microcavities with fixed layer thicknesses is limited considering a typical refractive index contrast of dielectric materials. In contrast, in the chirped microcavity approach a broad stopband with a spectral width of over 600 nm with a center wavelength around 800 nm can be achieved as depicted in Fig. 1b. The displayed reflectivity spectrum was computed using the transfer matrix method, considering the parameters specified in the supplementary information (SI) Table S1. In this example, the starting wavelength value $\lambda_1$ is 500 nm, and the design wavelength separation ($\Delta\lambda = \lambda_i - \lambda_{i+1}$) between adjacent wavelengths is 50 nm, with the final value of $\lambda_9$ being 1050 nm. $\Delta\lambda$ and the design wavelengths can be modified based on requirements of the active material. Additionally, by increasing the number of pairs for each wavelength step we can achieve an overall reflectivity close to 100%. In this calculation study, four pairs are considered for each design wavelength. After determining the design parameters, $SiO_2$ and $Si_3N_4$ layers with specific thickness are deposited on Si substrate by employing plasma-enhanced chemical vapor deposition (PECVD). The DBR layers are deposited at a temperature and pressure of 130 °C and 4 Pa, respectively, along with a plasma generator power of 50 W. These fabrication details, in combination with the chirped design, resulted in the successful realization of a microcavity with a broad stopband, offering new possibilities for advanced light-matter interactions.

### Chirped microcavity design

A chirped microcavity can be constructed by sandwiching a cavity spacer between two cDBRs. Unlike conventional microcavity configurations with a single resonance, the cDBR microcavity can generate multiple cavity resonances even at smaller effective cavity spacer thicknesses. This is due to the varying layer thickness, which fulfills multiple Bragg conditions over a larger spectral range. Normally, during the fabrication, the spectral position of cavity resonances can be adjusted by altering the thickness of the cavity spacer. However, in chirped microcavities, the spectral position can also be affected by changing the direction of the thickness gradient. We calculate the reflection response for two distinct microcavity configurations, each with opposite thickness gradients. See Figure S1 and Table S2 in SI for additional details about the effect of thickness gradient direction and associated DBR parameters. In fact, the spectral position and Q-factor of the resonances experience significant changes with variations in the direction of the thickness gradient. Additionally, design flexibility in terms of altering refractive index choices per section, the possibility to adjust Q factors by individually varying repetitions of layer pairs for specific wavelengths or/and wavelength, can enrich microcavity features and add unrivalled application advantages. The strategic choice of chirp direction and symmetry plays a crucial role in determining the spectral distribution of cavity resonances. This level of control facilitates the design of monolithic chirped microcavities tailored for specific active media offers an extraordinary and novel approach of tailoring light-matter interaction experiments.

**Table 1: Design parameters for the chirped microcavity employed for HS multi-resonance experiments.**

| DBR parameters | | Bottom DBR | Top DBR |
|---|---|---|---|
| Material and refractive index | $SiO_2$ ($n_1$)<br>$Si_3N_4$ ($n_2$) | 1.4704<br>2.011 | |
| Layer thickness | | $\lambda/4$ | $\lambda/4$ |
| Total wavelength steps $m$ | | 9 | 8 |
| Starting wavelength $\lambda_1$ | | 950 nm | 920 nm |
| Wavelength separation $\Delta\lambda$ | | 80 nm | 50 nm |
| Pair repetition for each $\lambda$ ($N_\lambda$) | | 3 | 1 |

## Multi-mode coupling with excitons from TMDC heterostructure

We take advantage of the multi resonance generation in cDBR microcavity configurations for effectively coupling the energetically separated exciton resonances from an active medium to the cavity resonances. Chirped microcavities are ideally suited to access and enhance exciton resonances in TMDC MLs and HSs. Specifically, it is very appealing and practical to use such extraordinary microcavities for the incorporation of TMDC HSs with type II band alignment, since those structures present several spectrally distinct exciton resonances. For this purpose, the cDBR microcavity design mentioned above can be modified to address such individual exciton resonances that exhibit twist-angle-dependent spectral modifications. Here, we carefully choose the wavelength separation for top and bottom DBR differently to particularly arrange the spectral positions of the cavity modes with respect to theTMDC HS exciton resonances.

The cDBR microcavity design parameters to induce light-matter coupling with a TMDC HS are presented in Table 1. The optimized cavity structure consists of three layer pairs for each design wavelength for the bottom cDBR, and only one layer pair for the top cDBR, mainly for directional out-coupling and facilitated top-side optical accessibility. Figure 1c displays the calculated and measured reflection response of the microcavity structure with the observed modes labeled 1 to 9 over the wavelength range of interest. The highlighted areas in Fig. 1c illustrate the resonance wavelength range of the excitonic emission from different bare TMDC MLs at RT, as well as the emission of $IX$ from TMDC HSs ($WSe_2$/$MoSe_2$). The spectral location of $IX$ is determined at low temperature because of commonly-occurring insufficient PL intensity at RT. Remarkably, the spectral positions of the calculated and measured cavity modes show excellent agreement (cf. Fig. 1d). Therefore, we demonstrate that chirped microcavities based on dielectric materials can be designed and fabricated with precise control over the spectral position of the multiple cavity modes.

To investigate the multi-resonance coupling, we utilize mechanical exfoliation[34] and dry-transfer[35] techniques to prepare a $WSe_2$/$MoSe_2$ HS. We cleave the TMDC crystals with low adhesive tape and subsequently exfoliate on a PDMS gel strip to aid in the dry-transfer technique. After identifying and choosing appropriately sized MLs of $WSe_2$ and $MoSe_2$ under an optical microscope, the MLs are aligned with respect to the edge of each layer. Finally, we transfer the monolayers onto a cDBR while maintaining a substrate temperature of around 60 °C along with few layer hBN on top[36].

Figure 2a shows an optical image of the stacked HS with outlined $WSe_2$ and $MoSe_2$ ML regions in blue and red, respectively, with the hBN cap in black. The $WSe_2$/$MoSe_2$ HS system typically hosts the $IX$ species, for which its electrons and holes are located in the conduction (CB) and valence bands (VB) of the individual MLs[6,37] in conjunction with their $X$ resonances, as sketched in Fig. 2b. Figure 2c shows the RT PL emission from both MLs' $X$ , $WSe_2$ (blue) and $MoSe_2$ (red), along with that from the HS region (black curve). The twist angle of the fabricated HS is determined by measuring the second harmonic generation (SHG) at RT[38]. Therefore, the intensity as a function of the irradiated laser's polarization is recorded from the individual MLs on the substrate by employing a linearly polarized picosecond mode-locked laser. Accordingly, the $WSe_2$/$MoSe_2$ HS system's twist angle is determined to be 10° ± 1°, as detailed in section 4 of the SI. A corresponding schematic of the $WSe_2$/$MoSe_2$ HS system on cDBR substrate indicates the stacking order of the constituent MLs (Fig. 2d).

The extracted exciton emission wavelengths are listed in the SI (Table S3). Noteworthy, the observed spectral redshift for the $X_{WSe_2}$ within the HS is attributed to a change in the dielectric environment of the HS region[20]. However, due to the substantial twist angle of 10°, the constituent MLs of the HS exhibit weak interlayer coupling, resulting in the absence of $IX$ emission at RT. Nevertheless, $IX$ emission is observed at cryogenic temperatures with an emission wavelength of approximately 905 nm (refer to SI section 6 for further details).

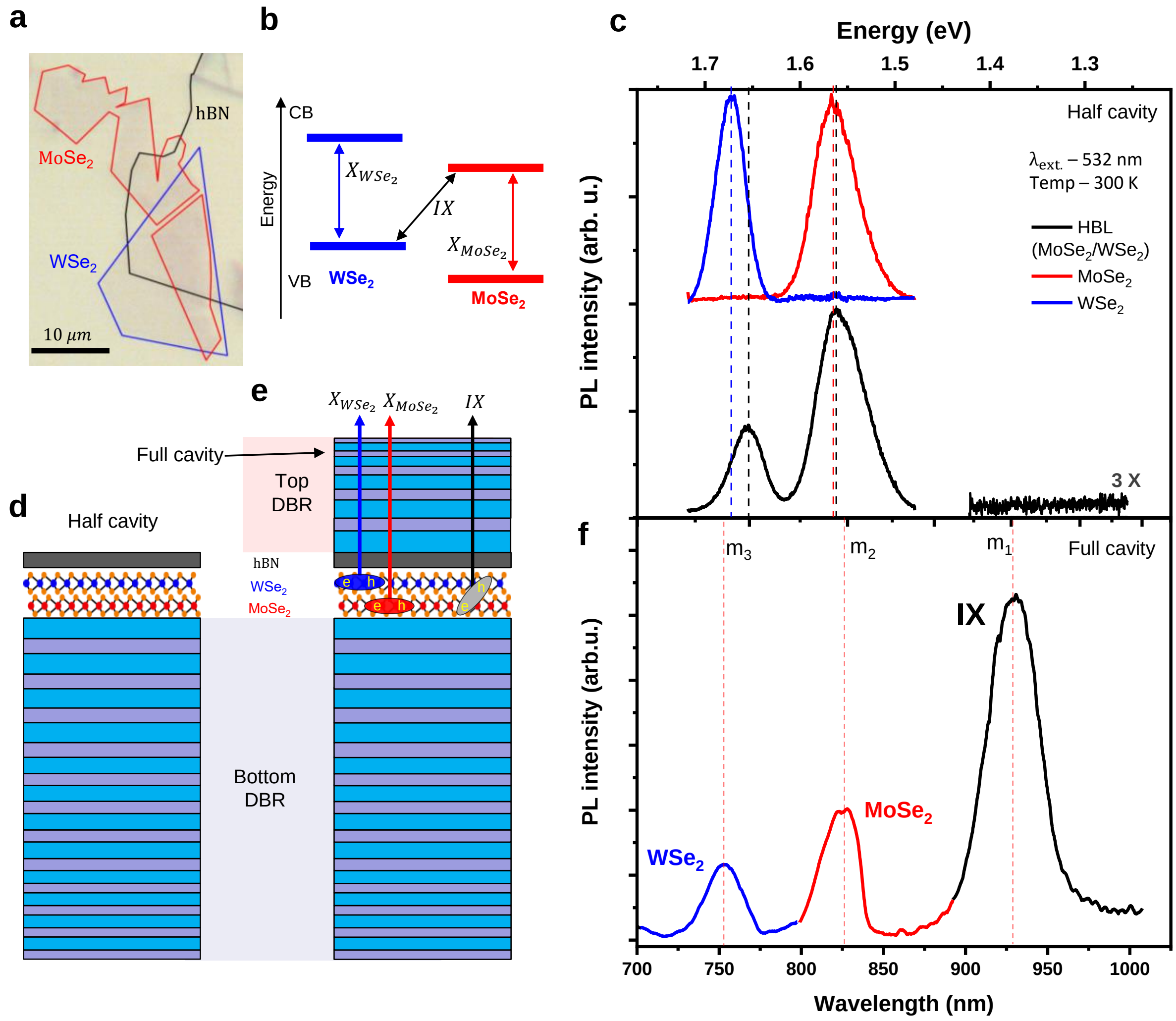

***Figure 2: Multi-mode Coupling with Excitons from a TMDC Heterostructure. a)*** *Optical micrograph of the fabricated TMDC ML HS system capped by hBN.* ***b)*** *Sketch of the type-II band alignment for the* $WSe_2/MoSe_2$ *HS spectrally distinct excitonic transitions (*$X_{WSe_2}$, $X_{MoSe_2}$ *and* $IX$*).* ***c)*** *Room temperature PL spectra showing* $X$ *emission from the* $MoSe_2$ *and* $WSe_2$ *MLs (red and blue curves, respectively) and the HS region (black). The* $IX$ *emission is not observed at RT. The dashed vertical lines indicate the extracted emission energy of the respective* $X$*.* ***d)*** *Schematic sample representation illustrating the HS's composition on top of a cDBR. The stacked HS possess twist angle of 10° ± 1° (cf. Fig. S2) and is capped with multilayer hBN.* ***e)*** *Analog to (d), the completed chirped microcavity system comprising the cDBR-supported HS covered by top cDBR layers. PL out-coupling indicated by color-coded upwards arrows.* ***f)*** *For direct comparison to (c), corresponding 300 K PL emission from the cavity-coupled HS system is plotted. Cavity mode positions* $m_1$, $m_2$, *and* $m_3$ *are indicated (dashed vertical lines). The system is excited with a continues wave laser at 532 nm.*

After the initial characterization of the fabricated TMDC HS, the top cDBR is deposited considering the parameters tabulated in Table 1. The thickness of the cavity spacer and the parameters of the top cDBR are carefully engineered to match the cavity modes spectrally with the relevant HS resonances (cf. Fig. 2). The results of a detailed investigation of the spectral position of the cavity mode with respect to the cavity spacer thickness can be found in the SI (Fig. S4).

Figure 2e shows the schematic of the fabricated microcavity system with integrated TMDC HS. We excite the HS system through the cavity mode, with a design wavelength of 532 nm aligned to the wavelength of our continuous wave laser. This scenario reduces the light absorption by cavity materials and leads to effective optical pumping of the HS system[39]. We observe three bright emission peaks, denoted as $m_1$, $m_2$, and $m_3$ in the PL spectrum measured for the chirped cavity at RT (Fig. 2f). This suggests that both $X$ and $IX$ couple with the corresponding cavity modes, thus in this experiment showing multi-mode coupling between excitons and cavity modes. The $m_2$ and $m_3$ modes couple with the $X$ resonance of $MoSe_2$ and $WSe_2$, respectively, while the $m_1$ mode couples with the $IX$. The emission wavelength of the coupled excitons along with full width half maximum (FWHM) is tabulated in Table S4 in SI.

Remarkably, we observe $IX$ coupling to the cavity mode $m_1$ in the form of a bright PL peak in the RT (300 K) spectrum recorded for this system, whereas such $IX$ signal is absent without the microcavity (see Fig. 2c). Note that, because the $IX$ emission obtainable from the HS system at low temperature without cavity is spectrally broad (cf. recorded 4 K spectra in Fig. S3), the $IX$ can significantly couple to the cavity mode at RT considering the temperature-dependent redshift of $IX$ [19,40,41] and the spectral position of $m_1$. This is evidenced experimentally here. The other two cavity modes, $m_2$ and $m_3$, are detuned by about 3 nm and 27 nm, respectively, in comparison to the $X_{WSe_2}$ and $X_{MoSe_2}$. However, due to the relatively large FWHM of emission (Table S3) and not too sharp optical modes (see Fig. S5 in SI), the MLs' $X$ emission, both $X_{WSe_2}$ and $X_{MoSe_2}$, satisfactorily couples to the microcavity. Next, we will discuss the impact of the cavity resonances on the PL yield of the HS system.

## Enhancement of interlayer exciton emission at room temperature

Cavity modes which are off-resonant compared to ML exciton resonances can indeed effectively suppress the spontaneous emission process for an exciton species. Oppositely, a mode matched to the wavelength of a specific emitter can Purcell-enhance its emission rate, such as the resonance designed to spectrally overlap with the $IX$ emission. Additionally, the cavity modes with a shorter wavelength, here $m_2$ and $m_3$ intended for the $X$, have an effective cavity length that are greater compared to the mode with a longer wavelength, that is $m_1$ for the $IX$. This implies that photons emitted by the $X$ experience a wavelength-dependent longer round-trip time than those arising from $IX$ (see Fig. S5a in SI) due to the specific cavity design and chirp direction. This effective difference between round-trip times leads to cavity modes with gradually increasing Q factors towards shorter wavelengths (see Fig S5b in SI). Hence, weaker or stronger light-matter coupling for $X$ emitters within the multi-resonant cavity can be designed individually (and differently) compared to the case for the $IX$ species.

In the multi-resonant cavity used, the coupling strengths vary for different species due to a possible detuning situation (cf. Fig. 2 RT spectra, $\Delta \approx 25$ meV for $X_{WSe_2}$) and the wavelengths-dependent Q factors. It is worth mentioning that, the charge transfer process responsible for $IX$ formation (on a timescale of < 100 fs) is much faster compared to the commonly observed $X$ lifetime on the order of ps [42,43]. On the other hand, the $X$ emission only partially couples to the assigned cavity mode $m_1$ due to spectral detuning at RT. As a result, the confined population of $X$ in both monolayers ultimately promotes $IX$ formation owing to the choice of the tailored cavity configuration. In contrast to the situation without a cavity, where RT signal is usually hardly obtainable[17–19], we observe an extraordinarily high intensity for the $IX$ emission even at 300 K. This is evidenced in $IX$ lifetime measurements discussed below which are enabled by the cDBR-induced intensity enhancement.

We also measured the temperature dependence of the cavity-coupled $IX$ emission (see Fig. S6 in SI). The recorded signals exhibit a maximum intensity at 20 K, indicating the optimum spectral overlap between the cavity mode and the $IX$ resonance. As expected, the PL intensity reduces drastically with increasing temperature due to thermal occupation of energetically higher $X$ states[44]. However, the $IX$ emission remains sufficiently coupled to the mode $m_1$ with FWHM of 15 nm, even with thermally-induced $IX$ redshift away from the design wavelengths. This is understandable because of the temperature-dependent spectral broadening of the $IX$, which maintains its coupling with that cavity mode. The results discussed here so far have presented an outstanding response after chirped cavity fabrication; nevertheless, these effects are even stronger evidenced through $IX$ lifetime measurements, as we discuss below.

## Temperature-dependent radiative lifetime of cavity-coupled interlayer exciton

Typically, the $IX$ species in such HS configuration exhibit nanoseconds long lifetimes owing to their indirect nature[5,42,45,46]. However, these excitonic transitions can be influenced by radiative coupling to the low-mode volume optical mode of a microcavity system, i.e. the Purcell effect. Moreover, the impact of temperature-dependent cavity–emitter detuning plays a role in lifetime modifications, which can be directly probed through time-resolved photoluminescence (TRPL) measurements. However, the $IX$ intensity reduces significantly with increasing temperature, making it challenging to systematically modulate the lifetime at elevated temperatures.

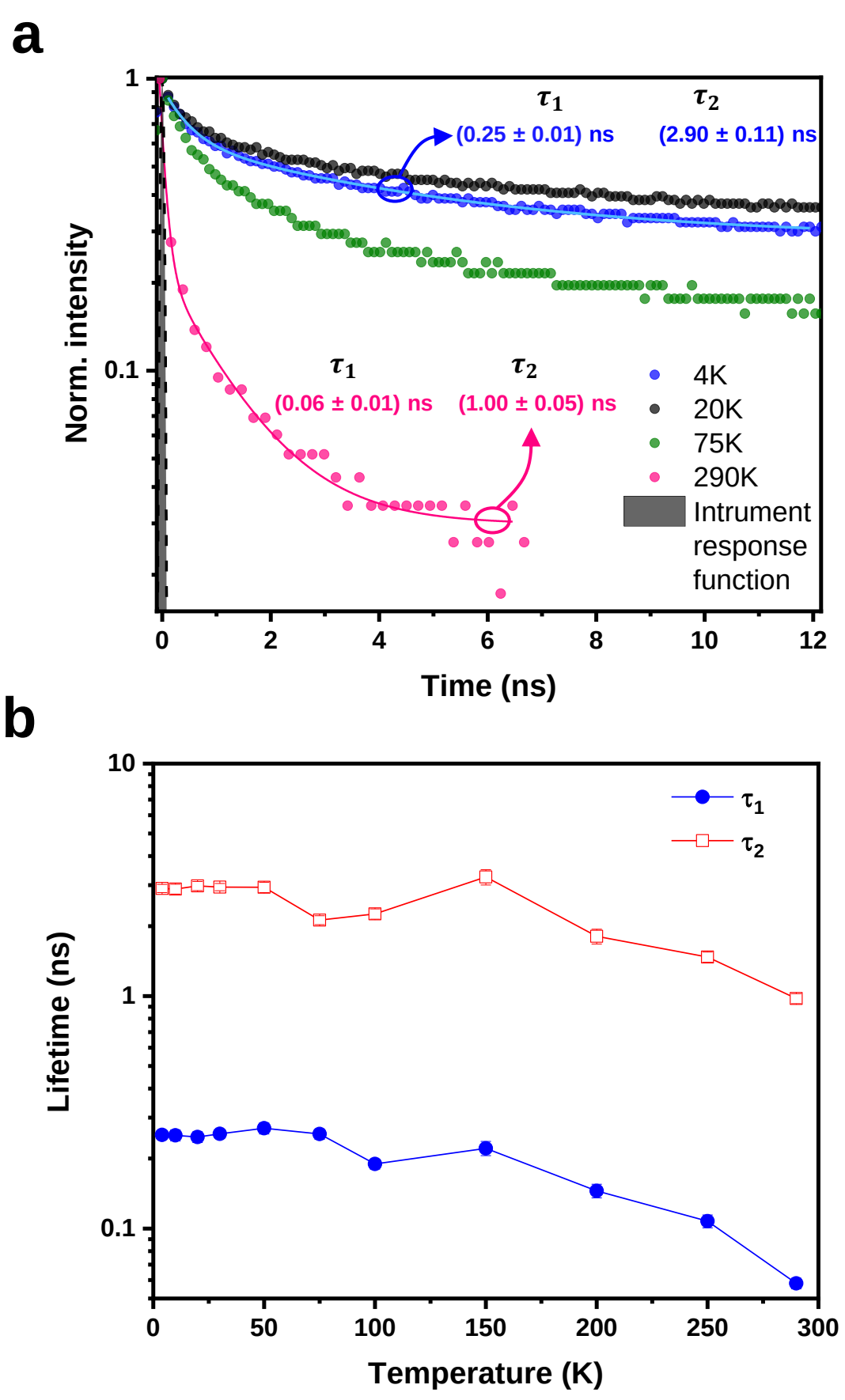

***Figure 3: Temperature dependence of interlayer exciton radiative lifetime**. a) Normalized PL intensity time trace of $IX$ signal in semi-logarithmic scale. The blue and pink solid lines show biexponential decay fits for time traces measured at 4 K and 290 K, respectively. The time trace constitutes fast ($\tau_1$) and slow ($\tau_2$) decay which are extracted through double exponential decay curve fitting.* ***b)*** *Measured lifetime components for the cavity-coupled interlayer exciton as a function of the temperature.*

Here, we utilize the cavity-enhanced emission intensity to investigate the temperature-dependent $IX$ lifetime. Figure 5a shows the TRPL-decay curves of the cavity-coupled $IX$ emission at various temperatures between 4 K and RT. We excite the HS region through the cavity mode with a picosecond pulsed laser at 741 nm, corresponding to the measured $WSe_2$ resonance (see Fig. 2c and Table S4). The collected PL signal is measured using superconducting nanowire single-photon detectors (SNSPDs). The instrument response function indicates sub-70-ps temporal resolution when measured using a SNSPDs. The recorded time-resolved emission decay of the cavity-coupled $IX$ is fitted with biexponential decay curves, constituting a fast and slow decay components, as plotted in Fig 3a. Commonly, the faster decay ($\tau_1$) is attributed to the bright $IX$ while the slower ($\tau_2$) is understood as a consequence of dark excitons and phonon-assisted decays[30,32,45]. Here, bright $IX$ are spin-allowed momentum direct (K-K) excitons, whereas the dark $IX$ arise from spin-forbidden transitions[30]. The measured traces support the attribution of two distinct time constants, but with a significant difference. At 4K, The cavity-coupled $IX$ exhibits fast (slow) decay time of (0.25 ± 0.01) ns ((2.90 ± 0.11) ns), which is considerably lower than reported values in the range of few-ns [30,32]. We attribute the observed reduction in radiative decay times to the Purcell effect by comparison with the intrinsic lifetime of (1.68 ± 0.49) ns from a bare HS sample of same configuration and with similar twist angle (see Fig. S7 in SI). Accordingly, our demonstrated Purcell factors are approximated to be 6.67 ± 1.02 and 2.60 ± 0.16 for the bright $IX$ species and the slower decaying $IX$ contributions, respectively. We attribute the smaller Purcell factor for the slow decay channel to the typically lower oscillator strength of optically dark species as well as phonon-assisted side-bands, which correspondingly exhibit less coupling to cavity modes. The evaluation of the Purcell factor is summarized in section 11 of the SI.

Intriguingly, the cavity-coupled HS configuration uniquely provides access to RT decay rate analysis for $IX$ emission. The perseverance of the cavity-emitter coupling evidenced by the enhanced PL signal is a strong proof of the effectiveness of the multi-resonant cavity system. That is, the chirped microcavity supports selective excitation and control of radiative emission channels for the relevant emitters both at cryogenic and at elevated temperatures. We observed a systematic reduction in decay times until 290 K (see Fig. 3b), which signify the coupling between the thermally-broadened and spectrally-shifted $IX$ and the cavity mode. The robust coupling between the cavity mode and $IX$ allows us to determine a fast decay rate as short as approximately (0.06 ± 0.01) ns at RT, together with a decay time of (1.00 ± 0.05) ns for the slow component. Thus, we demonstrate the Purcell enhancement of an $IX$ species not only in a so-far uniquely designed and novel multi-resonant microcavity configuration at cryogenic conditions, but also even up to RT. By that, our work underscores the utility of cavity-enhanced emission intensity in chirped microcavities as a powerful tool for investigating numerous exciton systems at elevated temperatures in general. The presented outcomes suggest a strong adaptability of our experimental approach, highlighted by the pronounced Purcell effect for *IX* emission in an archetypical TMDC HS. Our findings act as a foundation for further sophisticated experiments involving cavity-coupled HS systems with much more freedom with regards to accessible temperature ranges and emission rates.

## Conclusion:

We demonstrated a unique and innovative multi-resonant microcavity approach based on cDBRs. We highlight and exemplify the opportunities of this concept by enhancing the $IX$ emission from a cavity embedded TMDC HS up to room temperature. The effective emitter-field coupling in the Purcell regime for the multi-resonant system with three relevant spectrally separated exciton species was probed by temperature-dependent $IX$ emitter intensity and lifetime measurements. For this purpose, we designed and fabricated a novel microcavity configuration with cDBRs based on mirror pairs of $SiO_2$ and $Si_3N_4$. Such cDBRs exhibit an extraordinarily broad stopband width of more than 600 nm, thereby covering a wide spectral window from VIS to NIR wavelengths. In fact, as predicted by our numerical modelling and verified by experimental results, the fabricated microcavity structure hosts multiple optical cavity modes with adjustable spectral positions and Q factors by design. These modes then couple to the excitons in the integrated $WSe_2/MoSe_2$ heterostructures, as an experimentally determined $IX$ Purcell factor up to 6.67 ± 1.02 at 4 K demonstrates. Further, we specifically utilize cavity-enhanced emission of the $IX$ to investigate of temperature-dependent lifetimes, which have been hardly accessible at room temperature. By that, we measured remarkably short radiative lifetimes of 60 ps. Overall, our advanced microcavity approach provides additional and important possibilities to examine and employ multiple excitonic resonances in complex 2D quantum materials and other material systems. This offers new building blocks towards achieving novel light sources and paves the way for various applications, such as microlasers and single-photon sources, by enabling specific control and manipulation of excitonic interactions with multi-resonant cavity configurations.

## Acknowledgement:

Financial support by the Deutsche Forschungsgemeinschaft (DFG) within the Priority Program SPP 2244 “2DMP” project Re2974/26-1 (ID 443416027), project Re2974/21-1 (ID 410408989), project RA2841/5-1 (ID 403180436), and project RA2841/12-1 (ID 456700276) are gratefully acknowledged. ARI thanks Dr. A. Fuss in Giessen for stimulating discussions on related topics and former team member F. Wall for contributing an optimized TMM frame for 2D-materials cavity simulations.

## Conflict of Interest:

The authors declare no conflict of interest.

## Author Contributions:

CCP conducted the experiments with the help of BR, NH, C-WS, IL and AK-S under the guidance of SR. ARI and SR conceived the study on chirped microcavities and supervised the design and implementation. Structure calculation was performed by CCP. The TMDC sample preparation is done by CCP and BR. The data was evaluated, discussed and summarized in a manuscript through contributions by all authors.

# Supplementary Information

# Enhancement of interlayer exciton emission in a TMDC heterostructure via a multi-resonant chirped microresonator up to room temperature

Chirag C. Palekar[1 *], Barbara Rosa[1], Niels Heermeier[1], Ching-Wen Shih[1], Imad Limame[1],

Aris Koulas-Simos[1], Arash Rahimi-Iman[2], Stephan Reitzenstein[1 **]

[1]Institut für Festkörperphysik, Technische Universität Berlin, Hardenbergstrasse 36, 10623 Berlin, Germany

[2]I. Physikalisches Institut and Center for Materials Research, Justus-Liebig-Universität Gießen, Heinrich-Buff-Ring 16, 35392 Gießen, Germany

*c.palekar@tu.berlin.de **; stephan.reitzenstein@physik.tu-berlin.de

## 1. Design parameters for the chirped DBR

The parameters mentioned in the table S1 are used to model the reflection response employing the transfer matrix method (TMM). The calculated reflection spectrum is shown in Fig. 1b and exhibits spectral stopband exceeding a width of 600 nm. The reflection response of multilayered DBRs is calculated using the modified TMM-based simulation code as used in Ref. [1].

*Table S 1: Design parameters for chirped DBR shown in Fig. 1b.*

| DBR parameters | | Bottom DBR |
|---|---|---|
| Material and refractive index | $SiO_2$ ($n_1$)<br>$Si_3N_4$ ($n_2$) | 1.4704<br>2.011 |
| Layer thickness | | $\lambda/4$ |
| Total wavelength steps $m$ | | 9 |
| Starting wavelength $\lambda_1$ | | 1050 nm |
| Wavelength separation $\Delta\lambda$ | | 50 nm |
| Pair repetition for each $\lambda$ ($N_\lambda$) | | 4 |

## 2. Design parameters for the chirped DBR microcavity

The design parameters for fully fabricated microcavity consisting of top and bottom chirped DBRs (cDBRs) are tabulated in table S2. Here, the parameters for bottom and top cDBRs are chosen to be different in order to showcase the versatility of the cDBR approach.

*Table S 2: Design parameters for cDBR microcavity shown in Figure S1.*

| DBR parameters | | Bottom DBR | Top DBR |
|---|---|---|---|
| Material and refractive index | $SiO_2$ ($n_1$)<br>$Si_3N_4$ ($n_2$) | 1.4704<br>2.011 | |
| Layer thickness | | $\lambda/4$ | $\lambda/4$ |
| Total wavelength steps $m$ | | 9 | 9 |
| Starting wavelength $\lambda_1$ | | 950 nm | 920 nm |
| Wavelength separation $\Delta\lambda$ | | 50 nm | 50 nm |
| Pair repetition for each $\lambda$ ($N_\lambda$) | | 3 | 3 |

## 3. Effect of chirp direction on spectral density of cavity modes

The cDBRs can be deposited by gradually changing the layer thickness. The direction of thickness changes, i.e. the chirp direction, can be altered to manipulate the spectral density of the generated cavity modes. Two exemplary spectra illustrating the effect of chirp direction on density cavity modes are shown Fig. S1. Here, ascending and descending thickness gradient are referred as positive chirp and negative chirp respectively. The microcavity consisting bottom and top cDBR with positive chirp and negative chirp (see Fig. S1 a), respectively, exhibits dense to sparse resonance spectral distribution from lower to higher wavelengths as calculated reflection shown in Fig. S1 b. On the other hand, negative chirp to positive chirp results into sparse to dense resonance spectral distribution from lower to higher wavelengths (see Fig. S1 c,d).

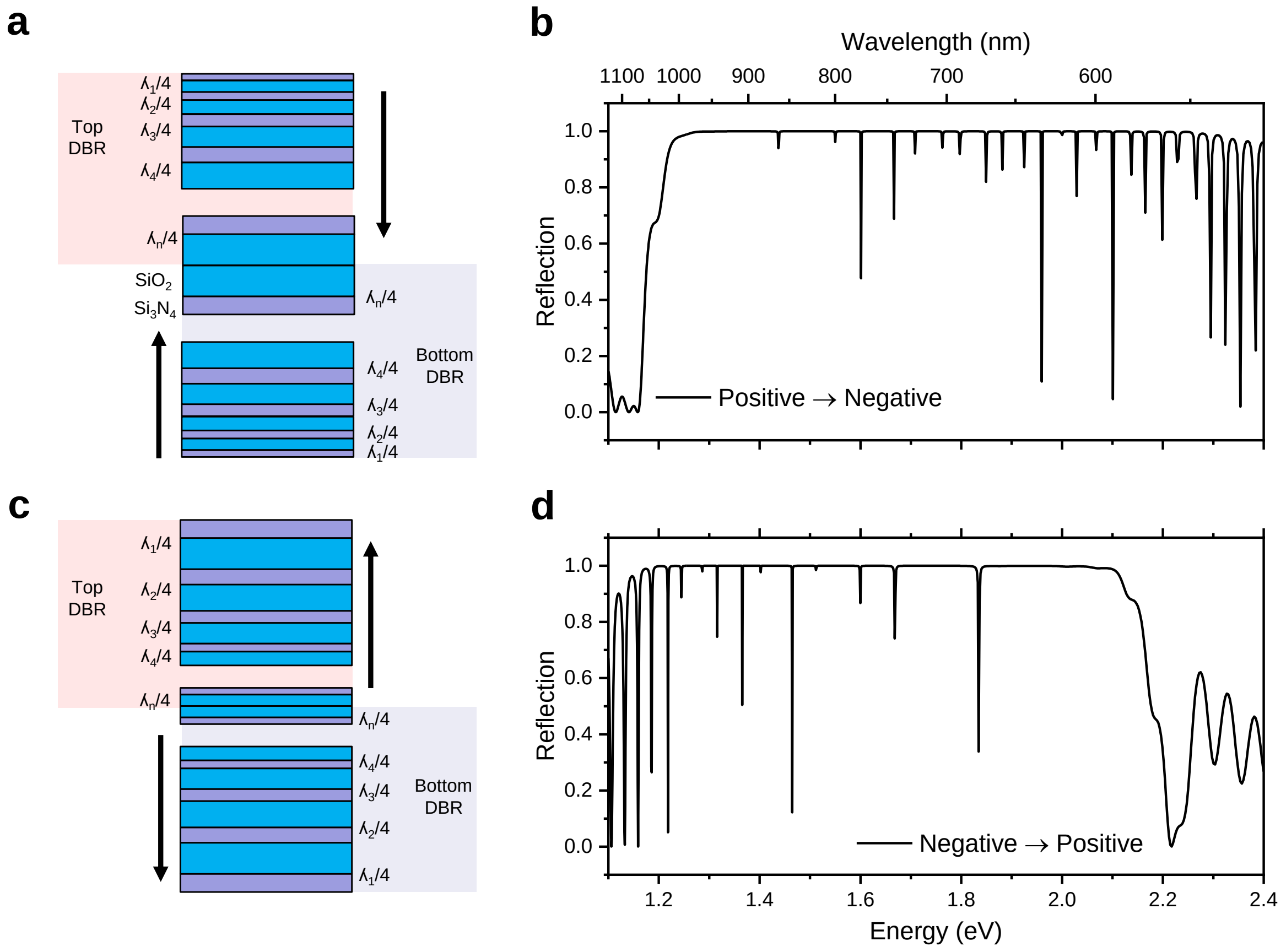

*Figure S 1: Influence of chirp direction of spectral density of resonances in a cDBR microcavity.*

### 4. Second harmonic generation for twist angle determination:

The second harmonic generation (SHG) from constituent MLs of the TMDC heterostructure allows us to determine the twist angle between the MLs. Here we excite the individual MLs with linearly polarized pulsed laser having excitation wavelength of 1313 nm and then we collect the SHG signal appearing at 651 nm with similar polarization. The measured SHG signal is fitted by a Gauss function for each polarization. Further, the SHG signal intensity from both MLs is recorded by systematically changing the laser polarization from 0° to 180° as shown in Fig. S2. Comparing the SHG response from the measured heterostructure gives a twist angle of 10° ± 1°.

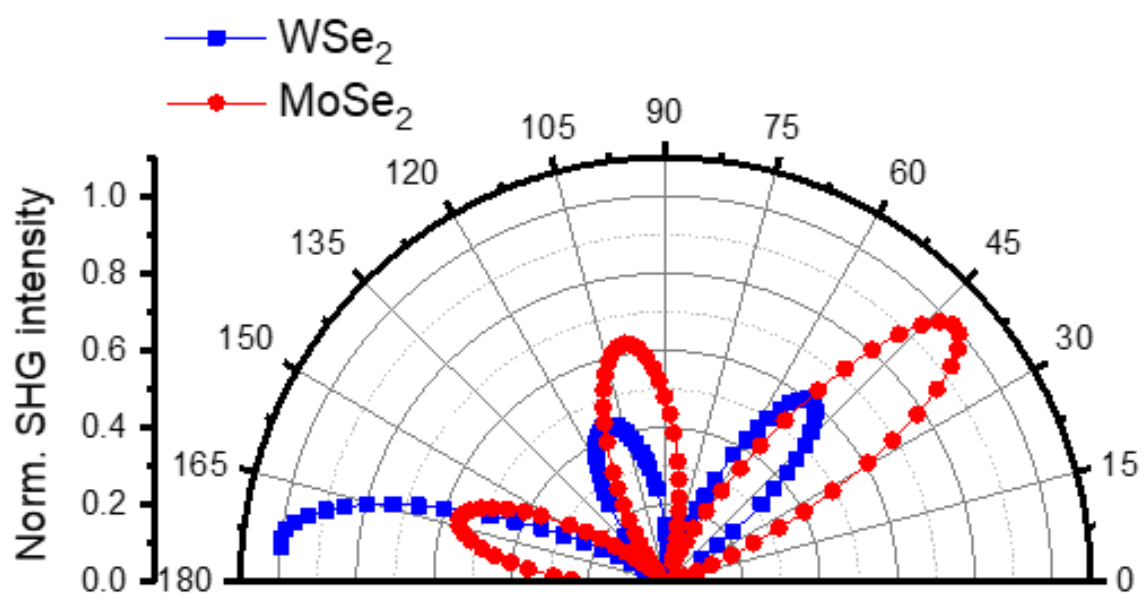

*Figure S 2: Polar plot of SHG response from the constituent ML measured at room temperature under pulsed laser excitation. Fitting the data yields a twist angle of 10° ± 1°.*

## 5. Photoluminescence emission of intralayer excitons in monolayer and heterostructure

The intralayer exciton photoluminescence emission from individual MLs and heterostructure is measured at room temperature is fitted with the Lorentz function to extract the emission wavelength and width (full width half maximum). The extracted parameters from the intralayer exciton emission are tabulated in Table S3.

*Table S 3: Extracted emission wavelength and width of the intralyer exciton emission from constituent ML and HS region at room temperature.*

| Region | Intralayer Exciton | Emission wavelength (nm) | Width (nm) |
|---|---|---|---|
| $WSe_2$ ML | $X_{WSe_2}$ | 741.3 | 20.1 |
| $MoSe_2$ ML | $X_{MoSe_2}$ | 793.6 | 26.9 |
| HS | $X_{WSe_2}$ | 749.0 | 20.3 |
| | $X_{MoSe_2}$ | 796.7 | 35.1 |

## 6. Interlayer exciton emission at cryogenic temperature

The interlayer exciton emission from the heterostructure region is absent at room temperature without a top DBR even at higher excitation laser powers. This is due to increased momentum mismatch and interlayer coupling between the layers caused by the relatively large twist angle of 10° in the present TMDC HS[2]. Figure S3 shows the interlayer exciton emission measured at cryogenic temperature of 4 K under pulsed laser excitation with wavelength of 720 nm. We observe the characteristic interlayer emission from type II band alignment system such as $WSe_2$/$MoSe_2$ heterostructures typically reported with emission wavelength between 880 nm to 950 nm[2,3]. The $IX$ emission is still observable at relatively low excitation powers which suggests even with large twist angle the interlayer coupling is strong enough for the formation of spatially indirect exciton at low temperature.

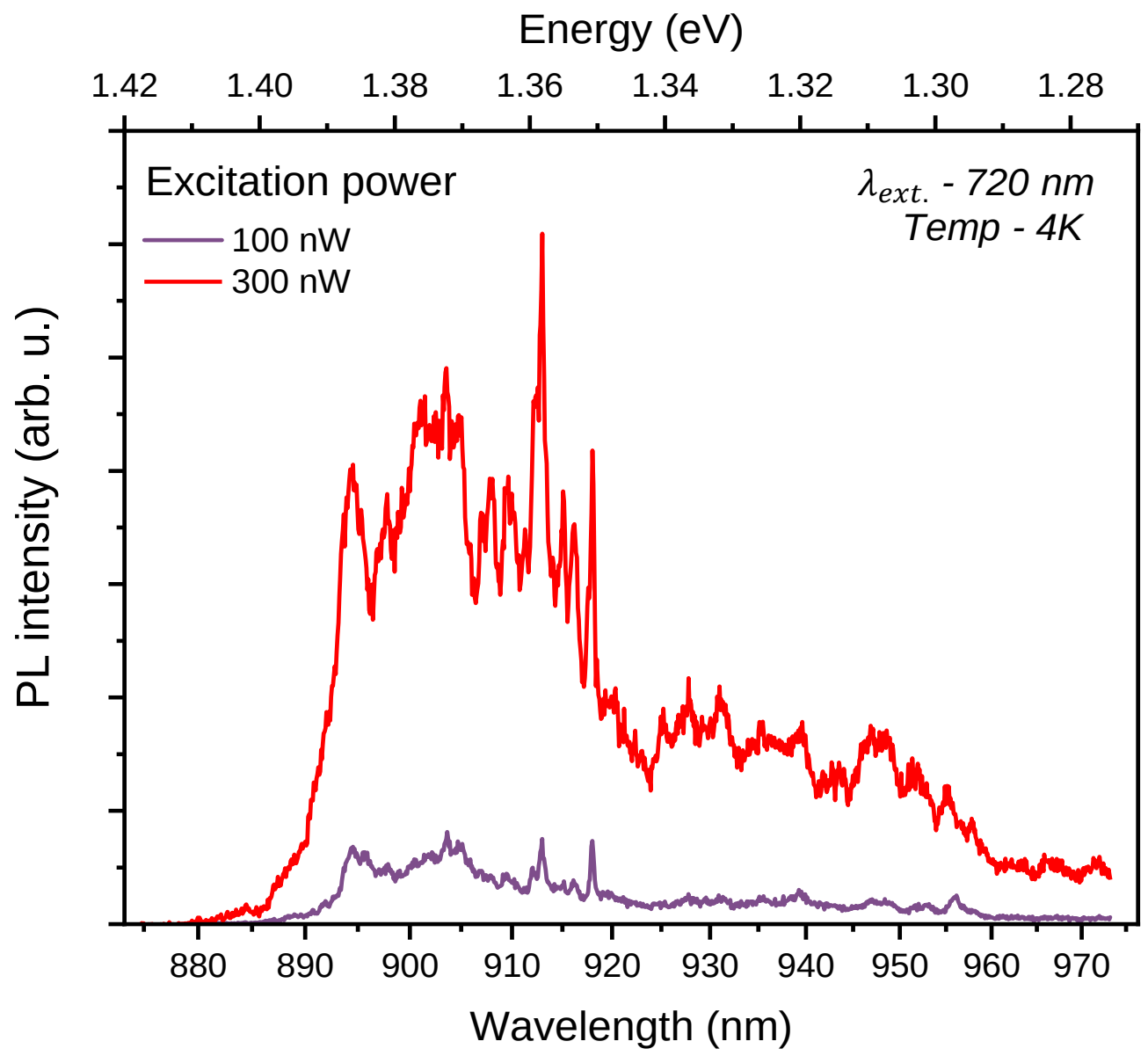

*Figure S 3: Interlayer exciton emission from the heterostructure region measured at low temperature.*

## 7. Effect of cavity spacer thickness on spectral position of cavity modes

We calculated the reflection response of the fully fabricated cavity as function of cavity spacer thickness. Here we utilize $SiO_2$ as a cavity spacer with thickness ranging from 100 nm to 1100 nm (see Fig. S4). By changing the cavity spacer length, the cavity modes can be tuned to the resonances of the active medium.

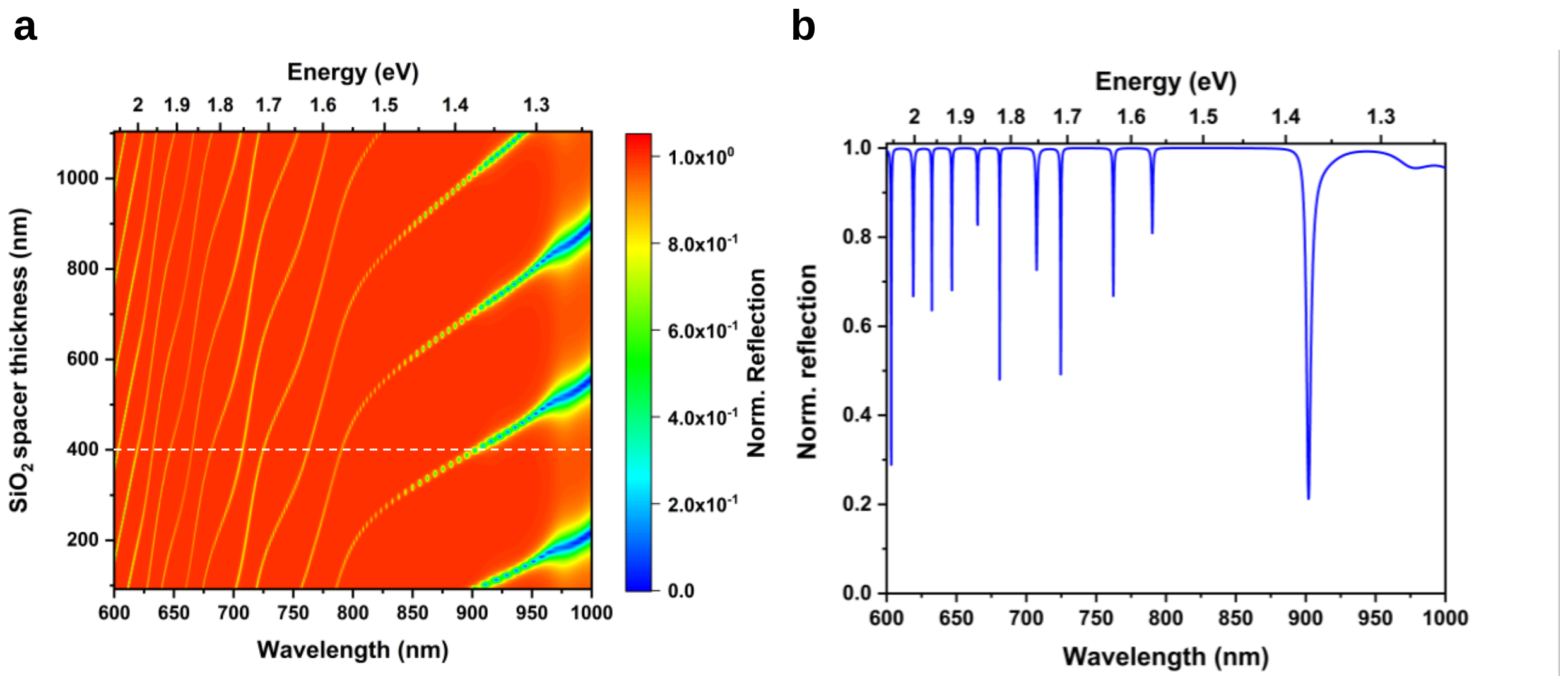

*Figure S 4: Calculated reflection response of the cDBR cavity with varying spacer thickness. a) False color map showing the influence of change in spacer thickness i.e. cavity length on spectral positions of the cavity modes. Calculated reflection of the microcavity system with cavity spacer of 400 nm also shown as dashed white line in a).*

## 8. Multi resonance coupling with excitons in heterostructure at room temperature

The intra- as well as inter-layer exciton from heterostructure couples to the cavity modes which demonstrates the multi resonance coupling at room temperature. We fit the observed emission from the cavity with a Gaussian lineshape function to extract the emission wavelength and width (FWHM). The extracted fitting parameters are tabulated in Table S4.

*Table S 4: Extracted emission wavelength and width of the intra and interlayer excitons from HS integrated in C-DBR microcavity at room temperature.*

| Region | Exciton coupled to cavity mode | Cavity mode | Emission wavelength (nm) | Width (nm) |
|---|---|---|---|---|
| $HS_{MoSe_2/WSe_2}$ | $X_{WSe_2}$ | $m_3$ | 752.9 | 25.5 |
| | $X_{MoSe_2}$ | $m_2$ | 823.5 | 38.4 |
| | $IX_{MoSe_2/WSe_2}$ | $m_1$ | 929.2 | 34.9 |

## 9. Variable effective cavity lengths and Q factor based on design wavelength

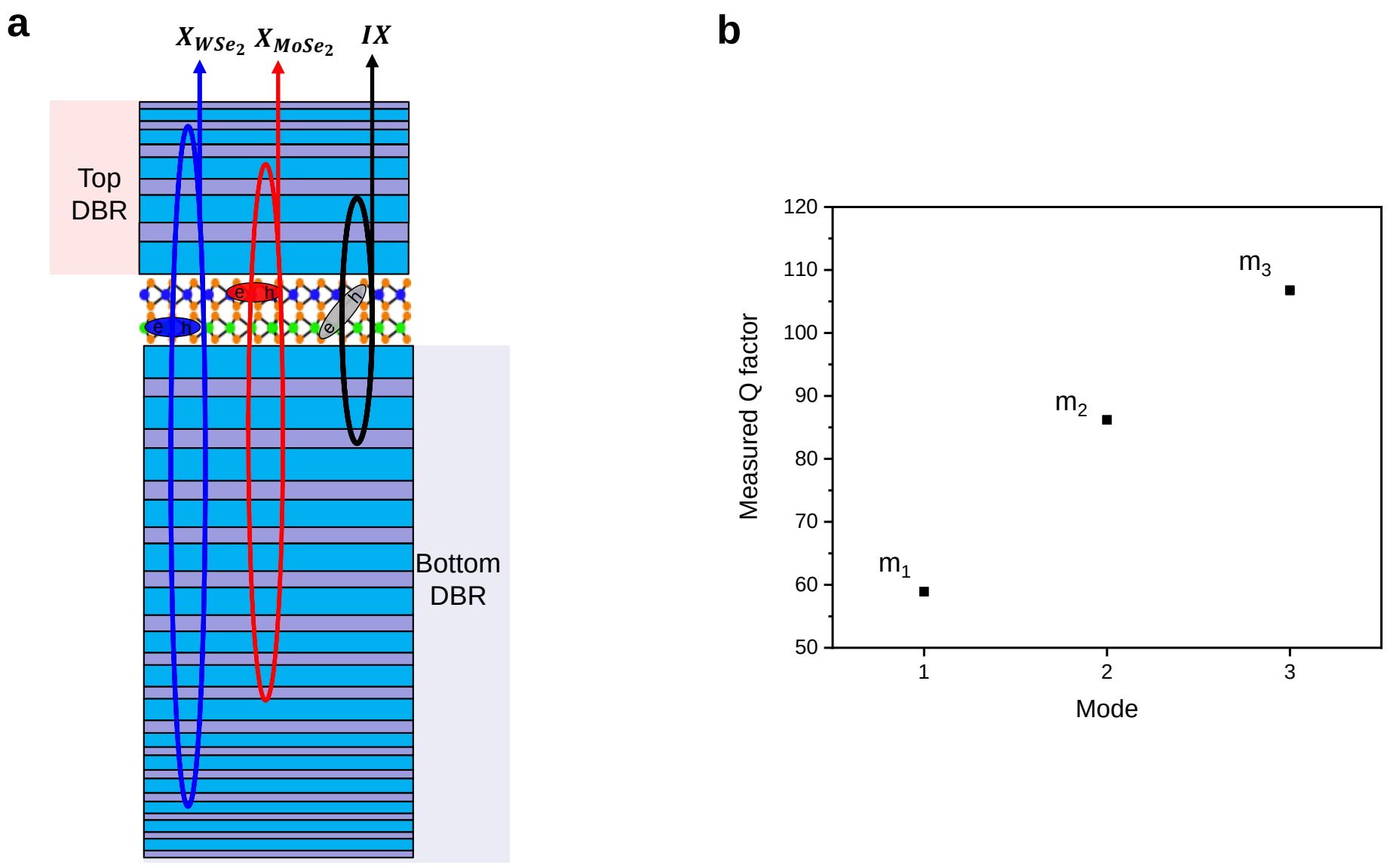

*Figure S 5: a) Illustration of the effective round trip path for photons with varying wavelength inside chirp cavity. b) Extracted Q factors of mode $m_1$, $m_2$ and $m_3$ from measured reflection response at room temperature.*

The effective cavity length for each design wavelength in cDBR depends on the chirp direction. A bottom DBR with positive chirp and a top DBR with negative chirp leads to effective cavity being larger for smaller wavelength and vice versa (see Fig. S5 a). This leads to increasing Q factors at lower wavelengths. The cavity mode ($m_2$ and $m_3$) coupling to the $X$ has relatively larger Q factor compared to cavity mode ($m_1$) coupling to the $IX$. Though enhancing resonance effects with higher Q factors, more spectral selectivity comes with it, too. Accordingly, tuning out of resonance towards elevated temperatures can lead to an inhibition of radiative recombination rate for the $X$ of such system.

## 10. Temperature dependence of cavity coupled interlayer exciton

The temperature dependent change is PL intensity is shown in Figure S6 along with extracted linewidth and mode energy. The PL intensity is measured for different temperatures in the range of 4 K to 290 K. The cavity coupled emission exhibit maximum intensity at 20 K indicating the optimum spectral overlap between the cavity mode and the $IX$ resonance. The emission intensity further reduces with increasing temperature due to detuning between the exciton resonance and cavity mode. Despite the increasing detuning at higher temperatures the coupling the cavity mode still persists due to the temperature-induced broadening of the exciton resonance (see Fig. S6 b).

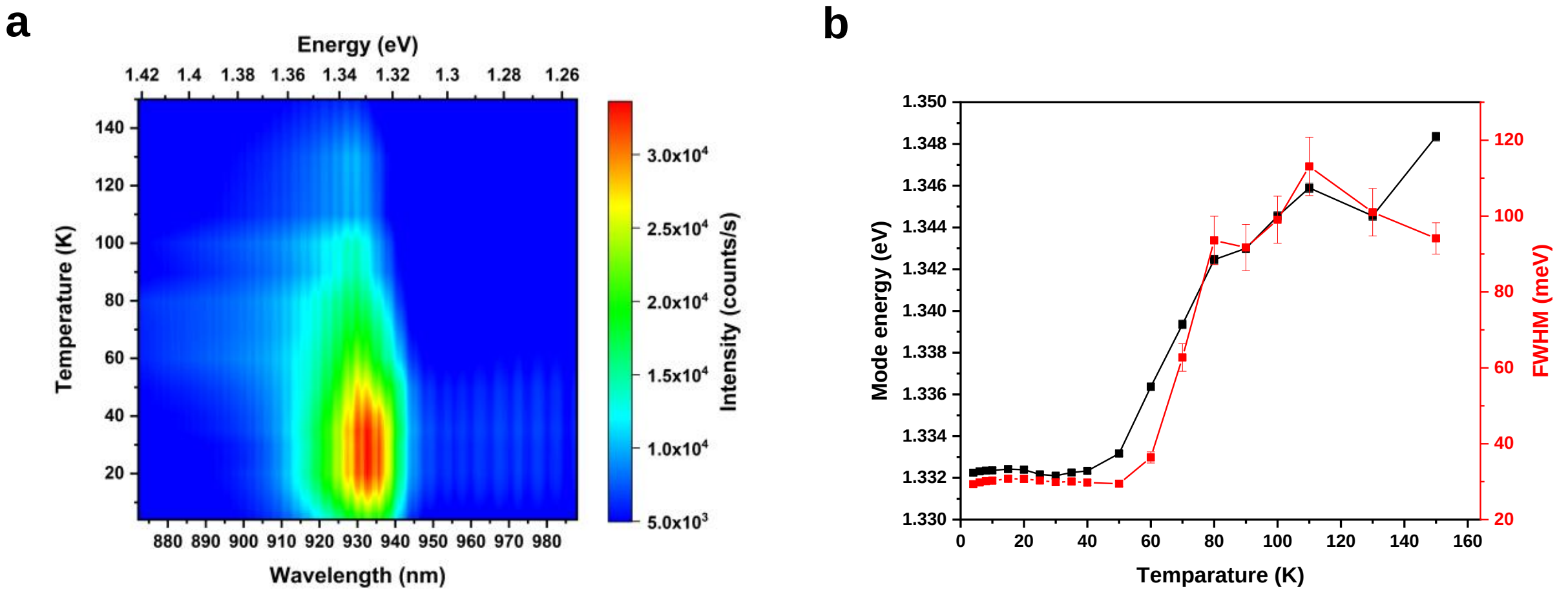

*Figure S 4: a) Interlayer exciton emission as function of temperature. The HS is excited with picosecond pulsed laser at 741 nm through the resonant cavity mode. b) Mode energy and FWHM of the mode coupled with the interlayer exciton exacted by means of a Lorentz fitting function.*

## 11. Purcell enhancement of interlayer exciton emission

In order to estimate the maximum Purcell enhancement, we performed time-resolved PL (TRPL) studies and compare the lifetimes extracted from the cavity-coupled $IX$ emission and from the reference $IX$ not coupled to a cavity from two separate nominally identical samples, both exhibiting a similar twist angle of 10°. The SHG measurements and the optical micrograph of the sample 2 are shown in Fig. S7 a.

Further, Fig. S7 b shows the measured emission decay from both samples at 4 K. The time trace from both samples constitutes biexponential decay. The cavity coupled emission shows faster decay rate (as shown in section: Temperature dependence of interlayer exciton radiative lifetimes) compared to intrinsic lifetime of $IX$ measured from HS region of sample 2. The $IX$ from sample 2 exhibits a fast (slower) decay time of (1.69 ± 0.49) ns ((7.55 ± 0.39) ns) at 4 K. Comparison of the TRPL results is shown in Fig. S7 b. We extract a Purcell factor of 6.67 ± 1.02 for faster decay and of 2.60 ± 0.16 for the slow decay channel of cavity coupled $IX$. The variation in Purcell factors is attributed to distinct coupling characteristics between the cavity mode and decay channels. The dark $IX$ have in-plane emission (orthogonal to the bright $IX$ and cavity mode) hence showcase lower coupling efficiency to the cavity mode[4].

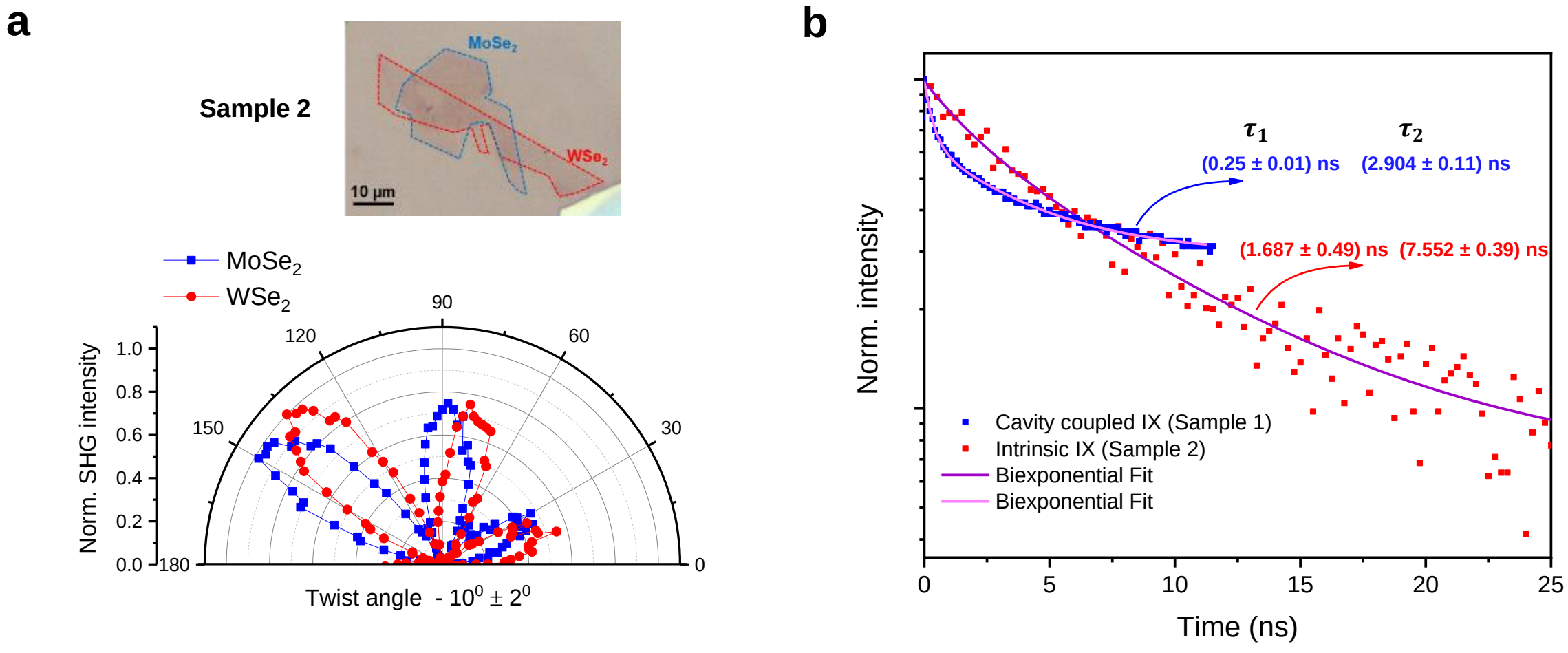

*Figure S 5: Purcell enhancement of interlayer exciton. a) Optical micrograph of Sample 2 and SHG response from individual MLs of $MoSe_2$ and $WSe_2$. Fitting yields a twist angle of 10° ± 2° for sample 2. b) TRPL data from sample 1 and sample 2 (blue and red, respectively) along with bi-exponential fits to determine the Purcell enhancement.*